\documentclass[a4paper]{mem}
\usepackage{natbib}
\usepackage{graphicx}
\usepackage{amssym}
\usepackage[a4paper]{hyperref}
\idline{73}{23}
\begin{document}
   \title{Numerical Results on Low Mass Star and Brown Dwarf Multiplicity
}

   \author{Eduardo Delgado-Donate \inst{1}, 
          \and
           Cathie Clarke \inst{2}\fnmsep
}

   \institute{Stockholm Observatory, AlbaNova University Centre, 106 91 Stockholm, Sweden \email{edelgado@astro.su.se}\\ 
              \and Institute of Astronomy, University of Cambridge, Madingley Road, Cambridge CB3 0HA, UK \email{cclarke@ast.cam.ac.uk}
             }

   \abstract{We have undertaken a series of hydrodynamic + N-body
   simulations in order to explore the properties of young stars. Our
   results suggest that the IMF may be sensitive to environment in its
   substellar region, with more brown dwarfs being formed where
   clusters are denser or more compact. We find that multiple stars
   are a natural outcome of collapsing turbulent flows, with a high
   incidence of $N > 2$ multiples. We find a positive correlation of
   multiplicity with primary mass but a companion frequency that
   decreases with age. Binary brown dwarfs are rarely formed, in
   conflict with observations. Brown dwarfs as companions are
   predominantly found orbiting binaries or triples at large
   separations.   } \authorrunning{E. Delgado \and
   C. Clarke} \titlerunning{VLMS Multiplicity Properties} \maketitle
%

\section{Introduction and Motivation}

   \begin{figure*}
   \centering
   \includegraphics[width=11cm]{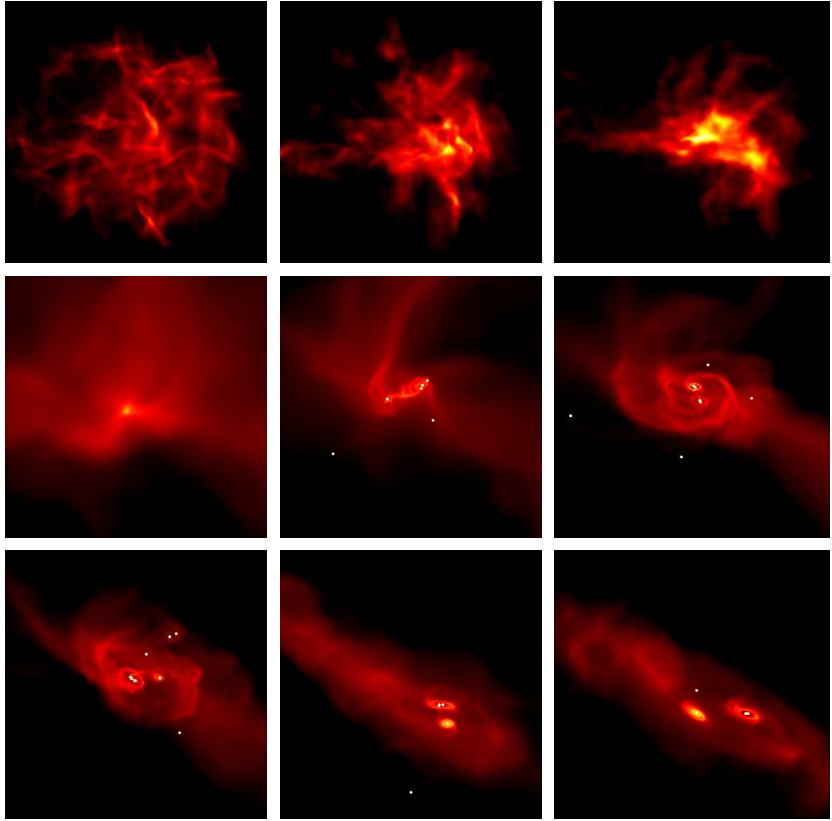}
   \caption{Evolution of a typical $\alpha=-3$ cloud}
    \end{figure*}

Most stars are known to be members of binary or even higher-order
multiple systems (Duquennoy \& Mayor 1991). Thus, any good star
formation theory must be a theory of (at least) binary star
formation. Currently we can hope to do more than produce multiple
stars by imposing some multi-armed instability on a collapsing
core. Turbulent initial conditions, for example, allow star formation
to be triggered in a less predictable way (e.g. Bate et al. 2003). In
addition, it has become computationally affordable to study the
statistics of star pairing beyond simple N-body integration
(Delgado-Donate et al. 2003). These two steps forward have made it
possible to perform calculations which both resolve the fragmentation
and collapse of molecular clouds (accounting fully for the
hydrodynamics) and which produce a statistically significant number of
stellar systems, thus opening the door to a direct comparison with
observations of the IMF and multiplicity properties of young stars
(Delgado-Donate et al. 2004a,b). In this paper we present the results
from the first hydrodynamic calculations to produce a statistically
significant number of {\it stable} multiple systems in the separation
range $1-1000$~AU. We will concentrate our attention mostly on those
aspects pertaining to the multiplicity properties of stars and brown
dwarfs.

\section{Numerical Scheme and Initial Conditions}

   \begin{figure*} 
   \centering 
   \includegraphics[width=13cm]{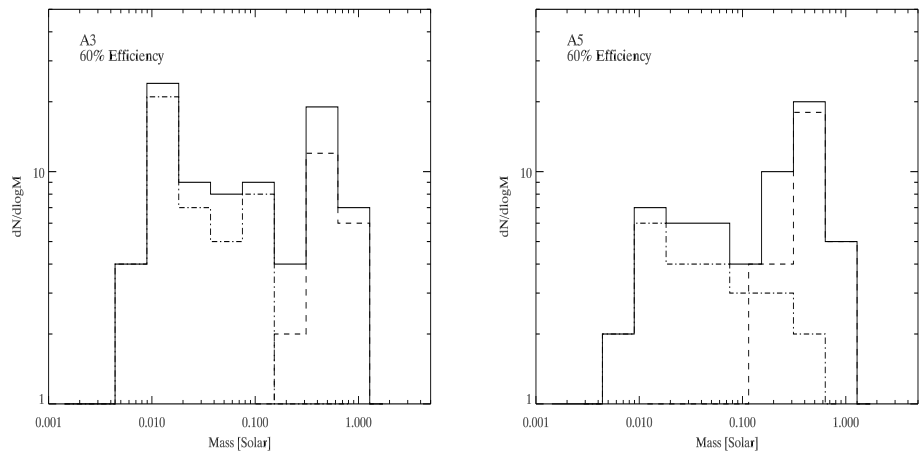}
   \caption{Mass Functions. Left, $\alpha=-3$; right
   $\alpha=-5$. The solid line includes all objects}
   \end{figure*}

We have performed 10 calculations of small fragmenting clouds, using
the SPH technique to solve the fluid equations. Our version of SPH
uses individual smoothing lengths and timesteps. Sink particles
replace bound blobs after a critical density is reached (Bate et
al. 1995). We apply standard viscosity with $\alpha=1$ and $\beta=2$,
and a binary tree to find nearest neighbours and calculate
self-gravity. The opacity limit for fragmentation (see e.g. Low \&
Lynden-Bell 1976) is modeled using a barotropic equation of state $p
\propto \rho^{\gamma}$, so that the gas is isothermal at low densities
($\leq 10^{-13}$~g~cm$^{-3}$) and polytropic with $\gamma = 5/3$ at
higher densities.

Each cloud is initially spherical, has radius of $\approx 10^4$~AU,
$5$~M$_\odot$ and density and temperature of $\approx
10^{-18}$~g~cm$^{-3}$ and $10$~K respectively. These values for $\rho$
and $T$ imply an initial Jeans mass of $\approx 0.5$~M$_\odot$,
typical of molecular clouds. We follow Bate \& Burkert (1997) and use
at least $100$~SPH particles to resolve the minimum Jeans mass that
can occur in the calculation (a few M$_{\rm J}$), thus resulting in a total
of $3.5 \times 10^{5}$~SPH particles.

We impose a random `turbulent' velocity field on each calculation,
defined by a power-law spectrum. The power-law exponent $\alpha$ is
set to $-5$ in 5 of the simulations and to $-3$ in the other half (the
former index corresponds to shifting the balance to having even more
power in large scales than in the $\alpha=-3$ case). These values of
$\alpha$ bracket the observed uncertainties in Larson's {\it velocity
dispersion-size} relationship. The velocity field is normalised so
that there is equipartition of kinetic and gravitational energy
initially. The velocity field is allowed to decay freely. We are
imposing a parameterised initial velocity field which approximately
reproduces observed bulk motions in molecular clouds (often described
as `turbulent' motions) but this term (`turbulence') should not be
taken to imply that we are modeling what a fluid dynamicist would
recognise as fully developed turbulence.

The gravitational force between sink particles is smoothed at short
distances. Therefore, binaries with semi-major axis $\lesssim 1$~AU
cannot form. A more detailed description of the code and the initial
conditions can be found in Delgado-Donate et al. (2004a).

The calculations are run until star formation no longer occurs (see
Figure~1 for snapshots of an $\alpha=-3$ calculation). This translates
into $\approx 0.5$~Myr of time or $60\%$ efficiency in terms of the
amount of gas converted into stars. At this point the remaining gas is
removed and the stellar system is evolved as a pure N-body system,
using NBODY2 (Aarseth 1999). After $10$~Myr we find that $95\%$ of the
multiples have decayed into stable configurations (using the criterion
by Eggleton \& Kiseleva 1995), and we stop the integration. The
calculations have been performed using the United Kingdom Astrophysical
Fluids Facility (UKAFF).

\section{Results: IMF}

Overall, the calculations produce $145$ stars and brown dwarfs after
$0.5$~Myr, an average of $\approx 15$ objects per calculation (a
number comparable to the initial number of Jeans masses). For the vast
majority of objects, the accretion rates are very low and so one
expects the objects masses to be representative of their {\it final}
masses. Overall, $40\%$ of the objects are brown dwarfs and $60\%$
stars. But the percentage changes depending on the initial value of
the turbulent spectrum slope. A higher fraction, nearly $50\%$ of the
objects in the $\alpha=-3$ calculations are brown dwarfs for $\approx
30\%$ in the $\alpha=-5$ case.

   \begin{figure*}
   \centering
   \includegraphics[width=5cm,angle=-90]{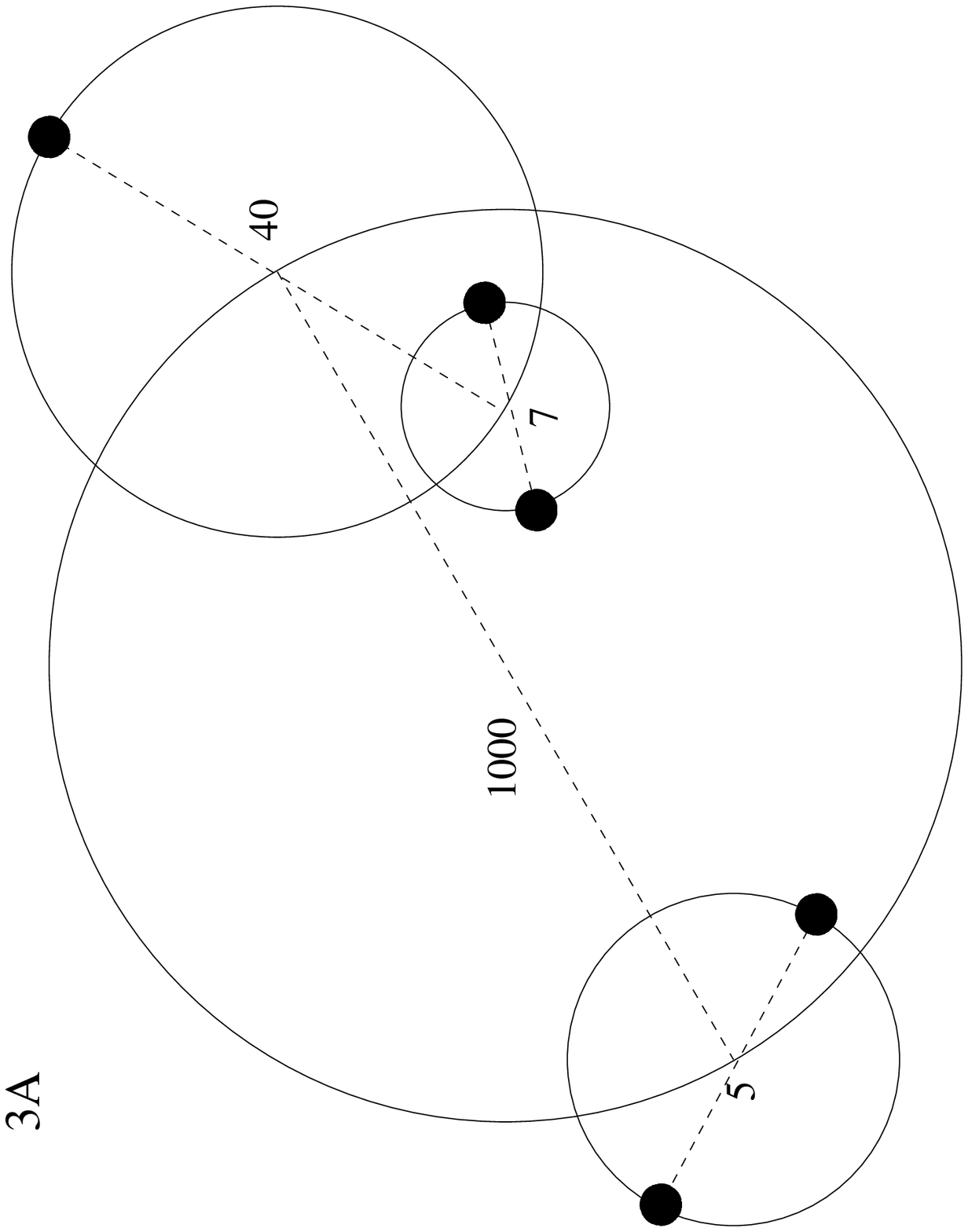}\includegraphics[width=5cm,angle=-90]{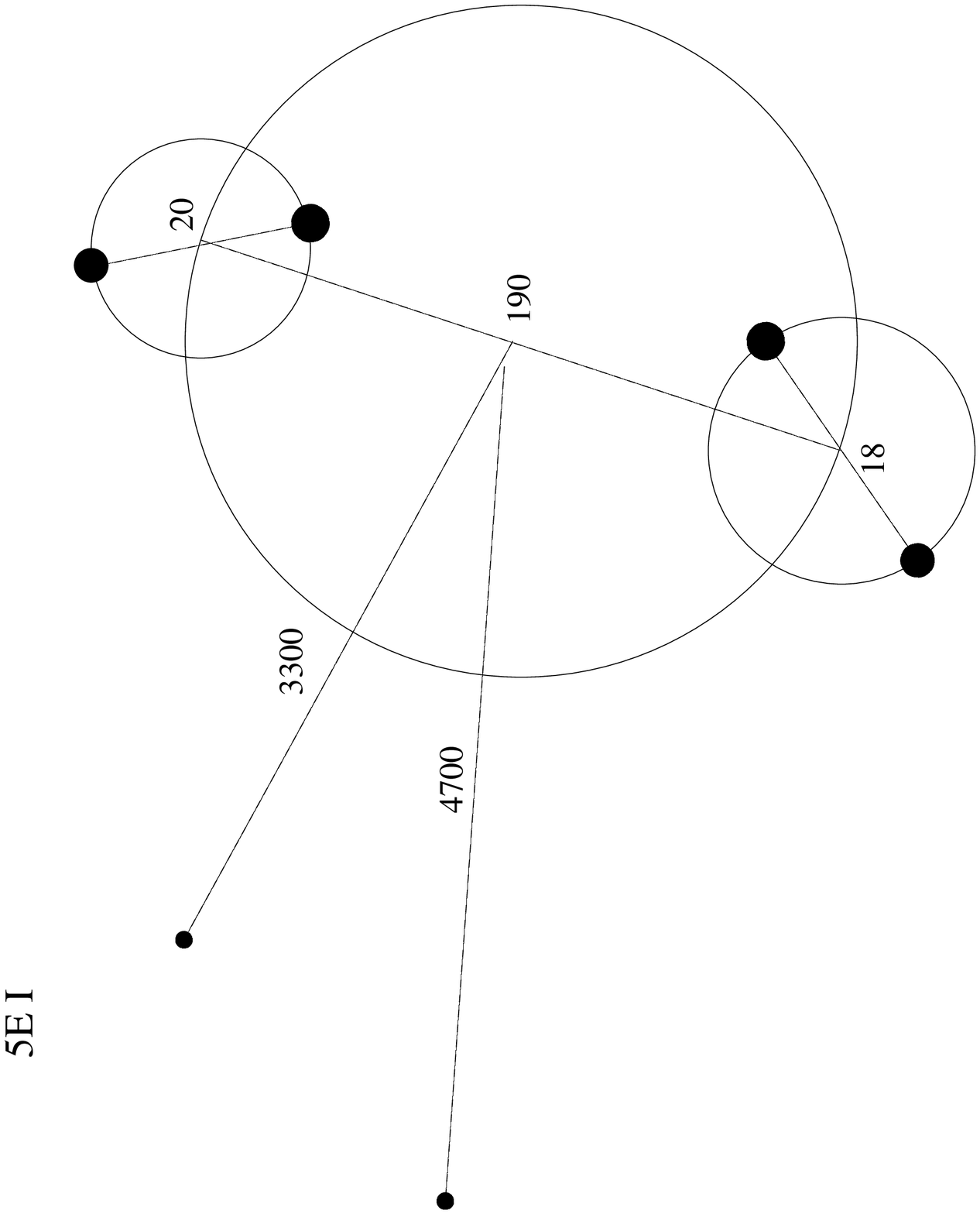}
   \caption{Examples of some of the {\it exotic} multiples that form in our calculations. Binary orbiting a triple and binary orbiting a binary plus some outliers.}
    \end{figure*}

The MFs derived from the two sets of calculations can be seen in
Figure~2. A KS test confirms that the MFs are different at the
substellar regime, at a $2\sigma$ level, while the stellar MFs are
indistinguishable. Notice that the upper-end of the MFs cannot be directly
compared with a Salpeter IMF as the MFs we present proceed from
calculations of clouds of the {\it same} mass, whereas the high-mass
end of the observed IMF is probably a reflection of the cloud mass
function, which is likely a power-law, non-flat distribution. It is
natural that the $\alpha=-3$ calculations produce a higher number of
brown dwarfs. These calculations have less kinetic energy in large
scales relative to the $\alpha=-5$ simulations and thus the action
tends to concentrate in dense, compact clusters near the cloud
centre. This leads to much more frequent dynamical interactions and
consequently ejection of low-mass objects, thus enhancing the fraction
of brown dwarfs. Other initial conditions (e.g. a denser cloud, see
Bate this volume) can also lead to a disparity in the fraction of
brown dwarfs formed. Thus, the substellar IMF appears as the region of
the IMF most sensitive to environment. Observational hints to this
conclusion can be found in the literature, e.g. Brice\~no et
al. (2002) find a paucity of brown dwarfs in Taurus relative to denser
SFRs like Orion (Slesnick et al. 2004).

\section {Results: Multiplicity Properties}

Our simulations produce a wealth of multiple systems. The multiplicity
fraction at $0.5$~Myr after the initiation of star formation is close
to $100\%$. The systems can adopt a variety of configurations, like
binaries orbiting binaries or triples (see Figure~3). It is apparent
that multiple star formation is a major channel for star formation in
turbulent flows.

\subsection{Multiplicity as function of primary mass}

The companion frequency decreases during the first few Myr of N-body
evolution, as many of the multiples are unstable. This internal decay
affects mostly low-mass outliers, which are released in vast amounts
to the field. We expect that in a real cluster the multiplicity would
drop even further as star forming cores do not form in isolation but
close to one another. Some of our binaries orbiting binaries might not
have survived in a more realistic environment. The predicted decrease
in the multiplicity frequency has been quantitatively observed by
Duch\^ene et al. (2004).
   \begin{figure*}
   \centering
   \includegraphics[width=8cm]{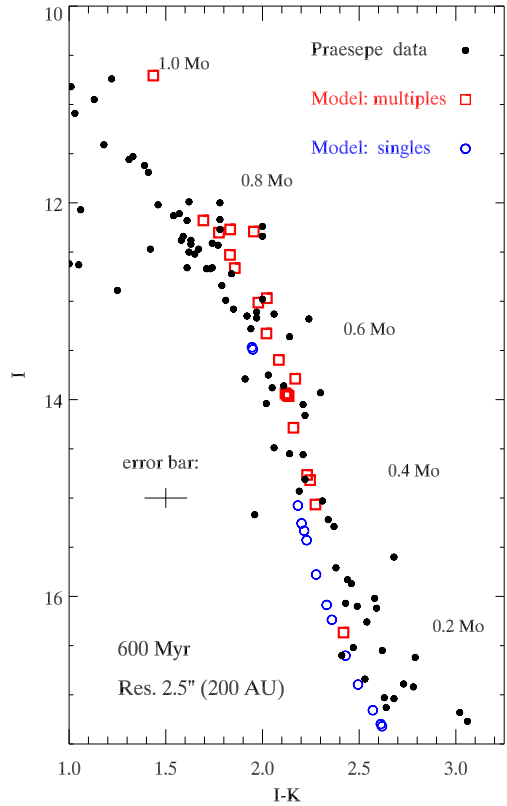}
   \caption{Colour-magnitude diagram ($I$ vs $I-K$) for the Praesepe clusters and superimposed our results. Symbols as in legend. Resolution at $200$~AU.}
    \end{figure*}

The properties of our multiples and the dependence of the binary
fraction on primary mass is best illustrated by a direct comparison
with the infrared colour-magnitude diagram of the 600 Myr old Praesepe
cluster (Figure~4). This comparison summarises the conclusions
obtained from a wider cross-check with observations. The cluster was
observed by Hodgkin et al. (1999) and our masses were converted to
magnitudes using the tracks by Baraffe et al. (1998). Binaries with
less than $200$~AU separation (the spatial resolution of Hodgkin et
al.'s measurements) are considered one object.

Two features from Figure~4 are worth noting: first, the simulated
cluster shows a binary sequence whose width is comparable to that of
the Praesepe, except for systems redder than $I-K=2.5$. This seems to
suggest that the formation of a significant minority of triples,
quadruples, etc. may indeed be common in real clusters. Second,
although our binary fraction for G stars is in agreement with
observations, our models fail to produce as many low-mass binaries as
observed. For example, at least a binary fraction of $15\%$ is seen
among brown dwarfs (e.g. Bouy et al. 2003).


\subsection{Where do we find brown dwarfs}

During the first few $\times 10^5$~yr most brown dwarfs are locked in
multiple systems, often orbiting a binary or triples in eccentric
orbits at large separations. Most of these systems are unstable and
decay in a few Myr, releasing individual brown dwarfs to the
field. Only a few substellar objects survive bound to stars. Of these,
the majority orbit a binary or triple at distances greater than
$100$~AU. One case out of 4 consists of a brown dwarf orbiting an
M~star at $10$~AU. Our results are in agreement with the observed
brown dwarf desert at very small separations (see e.g. Forveille, this
volume). However, more than a dozen substellar objects companion to
stars at wide separations are known (Gizis et al. 2001). According to
our results, we would expect that a large fraction of the primaries in
these wide systems should turn out, in closer examination, to be $N
\geq 2$ multiples.

\section{Conclusions}

We have undertaken the first hydrodynamical $+$ N-body simulations of
multiple star formation to produce a statistically significant number
of stable hierarchical multiple systems, with components separations
in the range $1-1000$~AU. We have shown that a high multiplicity
fraction is typical of the very early stages, a few $\times 10^5$~yr
after star formation begins, with many different possible multiple
configurations. At later stages, a few Myr, most systems have decayed,
ejecting brown dwarfs to the field and decreasing the companion
frequency. Both the high initial multiplicity and its dependence with
age seem to be in accord with resent observations. In addition we have
probed different power spectra for the initial random velocity field
and found that a larger fraction of brown dwarfs is produced when the
initial conditions favour a more compact, dense distribution of
stars. Our findings, taken together with others found in the
literature, seem to suggest that the substellar regime is where it is
most likely that the universality of the IMF might break down.\\

We find a positive dependence of multiplicity with primary mass, with
few low-mass stars being primaries. The paucity of brown dwarf
binaries in our simulations indicate that the models need finer
tuning. Brown dwarfs are found, however, orbiting binaries or triples
at large distances, and thus we suggest that a good test of our models
is to look into the primaries of wide brown dwarf companions in search
of multiplicity.

\begin{acknowledgements}
We gratefully acknowledge the Leverhulme Trust whose support (in the
form of a Philip Leverhulme Prize to CJC) allowed us both to attend
this meeting. EDD also wishes to thank the EU Network {\it Young
Clusters} and Svenska Vetenskapsr{\aa}det for their support.
\end{acknowledgements}

\bibliographystyle{aa}

\end{document}